\documentclass[a4paper,envcountsect]{llncs}
\usepackage[british]{babel}
\usepackage{amsmath,amsfonts,graphicx,multirow,url,algorithm,algpseudocode}
\usepackage{xspace}
\usepackage{xy}

\usepackage{graphicx}
\usepackage[font=small,skip=0pt]{caption}
\usepackage[export]{adjustbox}
\usepackage{url}
\usepackage{fancyvrb}

\usepackage[dvips,bookmarks=false,
            plainpages=false,naturalnames=true,
            colorlinks=true,
            linkcolor=blue,citecolor=blue,urlcolor=blue]
        {hyperref}

\usepackage{color}
\usepackage{fancyhdr}
\usepackage{url}

\newtheorem{theorem}{Theorem}[section]

\newtheorem{proposition}[theorem]{Proposition}

\newtheorem{assumption}[theorem]{Assumption}

\def\U{{\cal U}}

\def\ENC{{\cal{E}}}
\def\EMB{{\cal{I}}}

\def\NONSUBMIT

\def\U-ID{\text{\it {U-id}}}
\def\SP-ID{\text{\it{SP-id}}}
\def\AP-ID{\text{\it{AP-id}}}

\def\e-ID{\text{e-ID}}

\begin{document}
\raggedbottom
\title{Remote Document Encryption \\
        - \emph{\large encrypting data for e-passport holders} -}
\ifdefined\NONSUBMIT
\author{Eric R. Verheul\thanks{
Work done for the Dutch Vehicle Authority (RDW) on which a patent application has been filed.}}
\institute{
KeyControls, Radboud University Nijmegen,  \\
P.O. Box 9010, NL-6500 GL Nijmegen, The Netherlands.
\email{eric.verheul@[keycontrols.nl,cs.ru.nl]}\\
{\bf Version 9th June 2017}
}
\else
\author{}
\institute{}
\fi

\maketitle

\begin{abstract}
We show how any party can encrypt data for an e-passport holder such that only with physical possession of the e-passport decryption is possible.
The same is possible for electronic identity cards and driver licenses.
We also indicate possible applications.
Dutch passports allow for 160 bit security, theoretically giving sufficient security beyond the year 2079,
exceeding current good practice of 128 bit security.
We also introduce the notion of RDE Extraction PIN which effectively can provide the same security as a regular PIN.
Our results ironically suggest that carrying a passport when traveling abroad might violate export or import laws on strong cryptography.

\end{abstract}
\raggedbottom
\section{Introduction}\label{sec:intro}
Dutch citizens can access electronic government services through DigiD, a central authentication system. Citizens are redirected by the government service to DigiD where authentication takes place. If this is successful, the user directed back to the government service. Current DigiD is based on a knowledge authentication factor only (userid/password).
In 2016 a pilot was concluded supplementing DigiD with a possession factor.
This factor consists of a contactless electronic identity document of the user.
The user first authenticates to DigiD using userid/password and then lets DigiD remotely read his identity document whereby proving possession.
This setup is called {\em Remote Travel Document Authentication} \cite{RDA,RTDA}.
As the setup is also applicable to other types of documents we refer to {\em Remote Document Authentication} (RDA).
Compare Section \ref{sec:Basic_and_RDA}.
For simplicity of exposition, we only refer to electronic passports (e-passports) compliant with international specifications \cite{ICAO}.
Other identity document types are discussed in Section \ref{sec:extensions}.

In RDA one effectively uses an e-passport as a PKI smartcard to authenticate its holder to an external party.
Typically PKI smartcards also contain an encryption certificate allowing external parties to encrypt data for the user.
Gert Maneschijn posed the question if a similar functionality could be provided by an e-passport.
In other words: can an external party extract a public key from an e-passport allowing data encryption that can only be decrypted by the holder?
We refer to this ability as {\em Remote Document Encryption} (RDE). Note that we require that RDE is based
on existing e-passport protocols. The feasibility of RDE is not only academically interesting but also practically.
We discuss four practically relevant use cases but we think many more exist:
\vspace{-2mm}
\begin{description}
\item[Secure Messagebox] Governments are replacing paper letters with electronic messages made available to the user through a central facility. This facility is a hotspot, i.e. an accumulation of user personal data.
    RDE allows compartmentalization where various governmental organizations use user e-passports to encrypt data. In this way the central facility is no longer a hotspot.
\item[Medical Portal Compartmentalization]  Health care providers are giving patients electronic access to their medical records. For convenience medical portals are used allowing to perform both non-sensitive (e.g. making appointments) as sensitive functions (access to medical records).  Typically access is given to medical records of all departments the patient has consulted within the health care provider. RDE allows portal compartmentalization.
\item[Secure data storage on NFC devices] It is not good practice to locally store sensitive data in mobile applications (APPs) as these could compromise when the device is lost. This means sensitive data can only reside at server side and needs to be recollected when required. This can result in slow perceived APPs . The mobile data usage can also be costly. With RDE one can locally secure data. This is particularity interesting for Near Field Communication (NFC) enabled devices where e-passports can be directly read.
\item[Cloud encryption] A natural extension to the previous use case is to let a user connect a contactless card reader to a PC. The user can then RDE encrypt all its sensitive data and place it in the cloud.
    Only with the e-passport the data is decipherable again. In this setting users only need to (physically) protect their e-passport and no longer worry about cloud security.
\end{description}
Gert Maneschijn's idea was to base RDE on the e-passport AA protocol on which RDA is also based. That is, the AA public key would play the role of the RDE public key.
Specifically, if AA would be based on text-book RSA \cite{KATZ} one could base RDE on `forged' signatures.
Here a party can generate challenges enabling prediction of the signature.
The data would then be encrypted with the signature and the encrypted data would accompanied with the challenge. The e-passport holder would then be able to calculate the signature based on the challenge and decrypt the data. However, AA is not based on text-book RSA but is based on ISO/IEC 9796-2, Digital Signature scheme 1 \cite{9796-2}. Here the AA challenge is hashed by the e-passport and concatenated with a large random number generated by the document. This means it is not possible to predict signatures. Actually, many countries are no longer using AA based on RSA signature schemes but are replacing this with Elliptic Curve Cryptography (ECC) based signatures, i.e. EC-DSA \cite{ECDSA}. Here signature forgery is also precluded.

\vspace{2mm}
\noindent
{\em Outline of the paper}

\noindent
In this paper we show how RDE can be based on the Chip Authentication protocol.
Basic e-passport protocols and RDA are first discussed in Section \ref{sec:Basic_and_RDA}.
In Section \ref{sec:CA_and_RDE} we first present an RDE implementation metaphor and then discuss the Chip Authentication protocol on which RDE will be based.
RDE will be specified in Section \ref{sec:RDE_spec}. In Section \ref{sec:extensions} we further discuss RDE including some variations and extensions to identity cards and electronic driver licenses and
Section \ref{sec:conclusion} contains conclusions. Finally, in the appendix we provide results of our RDE experiments based on an e-passport simulator, a Dutch e-passport and a Dutch identity card.

\section{Basic e-passport protocols and RDA}\label{sec:Basic_and_RDA}
Like most card protocols, e-passport protocols distinguish the card and a party wanting to read data from it, usually called the {\em terminal}.
The terminal sends commands including parameters to the card called application protocol data units or APDUs to which the card responses and whereby certain states inside the card can change. Compare \cite{ISO/IEC7816-4}.
RDA \cite{RDA} is based on three basic e-passport protocols BAC, PA and AA \cite{ICAO11} we now briefly explain. All personal data printed on an e-passport is also electronically available. By the contactless nature of e-passports it needs to be avoided that this data can be surreptitiously read. To this end, access to the e-passport data is protected by {\em Basic Access Control} (BAC). In essence, the terminal needs to prove knowledge of a password. This BAC password takes the form of a so-called {\em Machine Readable Zone} (MRZ) a long string printed on the e-passport. As the name indicates, the MRZ is typically scanned by a computer facilitating electronic reading of electronic e-passports, e.g. in border control. Actually, only parts of the MRZ are actually used as a password, namely: the document number, the expiry date of the document and the date of birth of the holder. In BAC the e-passport generates a random number that needs to be encrypted by the terminal with a cryptographic key derived from the MRZ. Only when this encryption is correct, the e-passport will give the terminal access to its data. Actually, the BAC protocol not only ensures access control but also provides secure messaging protecting the communication against eavesdropping. This allows the terminal to protect the commands including the parameters and the responses. This secure messaging is based on two cryptographic keys, one for confidentiality and one for authenticity. These keys are based on the MRZ and random numbers generated by the card and the reader. With these keys, the terminal
encrypts the command (APDU) and parameters, calculates a message authentication code on the result and wrapped these in a secure messaging object. The card does a similar thing with the response. In \cite[Appendix D-4]{ICAO11} an worked out example of this is provided which is very illustrative.
From now on we will refer to the {\em protected command} and {\em protected response} in this context.

Data on an e-passport takes the form of (elementary) files on the document called {\em data groups}. These data groups are both individually and collectively signed by the government that issued the document. This is called {\em Passive Authentication} (PA). For PA an e-passport contains a file called the {\em Document Security
Object}, also known as $\text{EF.SO}_\text{D}$. The document security object contains a digital signature of the issuer and the public part of the key that was used. This public key is wrapped in a certificate issued from a published government Public Key Infrastructure. The signature is placed on the concatenation of the hashes of all data groups. Validation of a data group comprises of checking the concatenated hashes signature and checking that the hash of the data group is present in the document security object.

With PA one cannot only validate the authenticity of an individual data group but also that all data groups correspond to the same e-passport. As these data groups and signatures can be copied, e-passports can also prove their authenticity. This is called {\em Active Authentication} and works as follows. One data group (data group 15) contains an {\em Active Authentication} public key.  The corresponding private key is securely placed in the e-passport. A party reading the e-passport can validate is authenticity by generating a (8 byte) random number and letting this be signed by the e-passport with its AA private key. The party can validate the signature and the authenticity of the public key through PA.

In Protocol \ref{RDA} we have given a basic description of RDA \cite{RDA} of which many variants exist. The interlinking between the various steps is typically based on the Security Assertion Markup Language (SAML) protocol \cite{SAML}. Clearly, if the userid/password authentication in Step 3 fails one typically would allow the user to re-enter the userid/password for a limited number of times. The MRZ in Step 4 does not need to be provided to DigiD at each authentication but could be stored by DigiD, e.g. as part of a registration process.
 \begin{algorithm}
\floatname{algorithm}{Protocol}
\caption{\em Remote Documentation Authentication}\label{RDA}
  \begin{algorithmic}[1]
  \State \texttt{A service provider requests a user to authenticate.}
  \State \texttt{The user is redirected to DigiD.}
  \State \texttt{The user authenticates at DigiD with userid/password. If authentication is unsuccessful RDA fails.}
  \State \texttt{The user allows DigiD access to its e-passport through middleware, i.e. the e-passport application is selected.}
  \State \texttt{The user allows DigiD to read his e-passport by giving the MRZ. If unsuccessful RDA fails, otherwise secure messaging is established.}
  \State \texttt{DigiD reads data group 1 (Identity Information), data group 15 (AA public key) and the Document Security Object  ($\text{EF.SO}_\text{D}$).}
  \State \texttt{DigiD validates the authenticity of the data groups read and that they belong to each other (PA). If unsuccessful, RDA fails.}
  \State \texttt{If the identification data from steps 3 and 6 do not match, RDA fails.}
  \State \texttt{DigiD generates a random number and lets the e-passport sign this with its AA private key (AA).}
  \State \texttt{DigiD validates the signature. If this is incorrect, RDA fails.}
  \State \texttt{RDA is successful and the user is redirected to service provider accompanied with relevant personal data, e.g. the BSN.}
  \end{algorithmic}
\end{algorithm}
If Step 5 is successful the commands including parameters and responses in Steps 6 and 9 are protected. One can also use another protocol controlling access to the e-passport in Step 5. This protocol is called Password Authenticated Connection Establishment (PACE)\cite{ICAO11}. In essence, PACE is based on the establishment of a secure messaging tunnel based on an (anonymous) Diffie-Hellman key exchange based on the MRZ data. If a reader does not use the correct MRZ data, the tunnel will not be established and the e-passport data cannot be read. PACE is more secure than BAC as it is more resistant to MRZ brute-forcing. In Step 9 of the RDA protocol one can also incorporate a message, e.g. by letting the random number be based as the hash of a message and a random number. In this way, unrefutable proof exists that the user, or rather its e-passport, was present during the authentication and consented to the message. Actually, this property is the reason that German e-passports do not support the AA protocol as it could be abused. Compare \cite[Appendix C]{ICAO11}. Alternatively, one can use the Chip Authentication (CA) protocol which, similar to the PACE protocol, is based on the Diffie-Hellman key exchange protocol. With the CA protocol the reader can validate the authenticity of the e-passport but in a non-transferable way. That is, a transcript of a CA execution cannot be used to other people as a proof that the user's e-passport was present. This is due to the property that anyone can construct such CA transcripts. Actually the CA protocol has an even stronger property which forms the basis for our Remote Document Encryption setup as we show in the next section.
\section{RDE through chip authentication}\label{sec:CA_and_RDE}
\begin{figure}[H]
\includegraphics[scale=0.2,center]{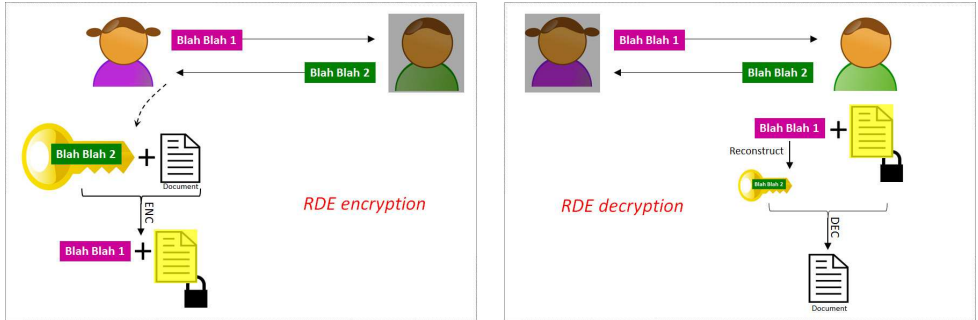}
\caption{RDE metaphor} \label{fig:RDE_metaphor}
\end{figure}

\begin{figure}[H]
\includegraphics[scale=0.3,center]{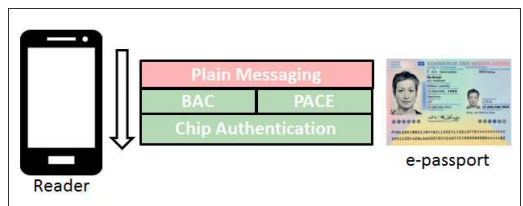}
\caption{e-passport secure messaging} \label{fig:SM}
\end{figure}

Consider two people, Alice and Bob, having regular conversations.
Alice knows Bob so well, that she knows at forehand the responses of Bob of any statement she makes.
Alice can then encrypt messages for Bob as follows (cf. Figure \ref{fig:RDE_metaphor}):
\vspace{-2mm}
\begin{enumerate}
\item Alice generates a random statement and predicts the response of Bob.
\item Alice uses the response as a key to encrypt a message for Bob.
\item Alice sends the encrypted message to Bob plus her statement.
\end{enumerate}
On receipt of the encrypted message and Alice's statement, Bob can then decrypt the message as follows:
\vspace{-2mm}
\begin{enumerate}
\item Bob forms its response to the statement of Alice.
\item Bob uses the response to decrypt the encrypted message.
\end{enumerate}
To make this setup secure, it is vital that only Alice can predict Bob's responses.
\vspace{-8mm}
\vspace{-8mm}
As in the metaphor, we let a party use a predicted response of the holder's e-passport as an encryption key for a message.
To make the responses only predictable for the party (and the user) we base RDE on the {\em secure messaging} mechanisms of e-passports.
Information between an e-passport and a reader is exchanged radiographically and secure messaging protects against attacks like eavesdropping.
More in particular, secure messaging protects the confidentiality and authenticity of exchanged information.
Within the context of e-passports three types of secure messaging exist: BAC, PACE, and CA.
As as indicated in Figure \ref{fig:SM}, CA and its secure messaging can only be established after a successful execution of either the BAC or PACE protocol.
After a successful execution of the CA protocol, BAC/PACE secure messaging is {\em replaced} by CA secure messaging.
Moreover, the establishment of CA secure messaging is independent of the BAC/PACE tunnel.
In other words, for the description of CA messaging we effectively do not have to consider BAC/PAC.
For completeness we have also mentioned the {\em Terminal Authentication} (TA) protocol. With this protocol, the terminal can prove its authenticity to the e-passport similar to the AA protocol. This is required before sensitive data groups (biometrics) can be read. TA is mentioned as ``Extended Access Control to Additional Biometrics'' in the ICAO specification \cite{ICAO11} but is not further specified. In practice, the TA protocol specified in \cite{BSI} is used. TA retains secure messaging from CA, i.e. does not introduce additional secure messaging.
The CA protocol works as follows. Similar to the AA setup the CA setup also contains a public key stored in an e-passport data group (data group 14, cf. \cite{ICAO11}). This data group is also signed by the publisher through passive authentication. The CA private key is securely stored in the document chip. In contrast to the AA setup, the CA public key is not a digital signature key but a so-called Diffie-Hellman key, cf. \cite{KATZ}. The CA protocol is based on an additive group $(\langle G \rangle, +)$  of order $q$ generated by an element $G$; these are called {\em domain parameters}. The CA public key $Y$ on an e-passport takes the form $xG$ where $x\in_R (0, q)$ is the CA private key. We use additive notation as this is customary in the context of elliptic curve groups which are now commonly used in CA. As an illustration, all active Dutch passports (and identity cards) are equipped with an elliptic curve CA key. Prior to 9 March 2014, Dutch passports used the 256 bit brainpoolP256r1 curve and after this date the 320 bit brainpoolP320r1 curve is used.

Data group 14 also contains the symmetric cryptographic algorithms supported in CA secure messaging.
This encompasses an encryption algorithm and a message authentication (cryptographic checksum) algorithm.
Secure messaging is based on encrypt-then-authenticate mode, i.e. data is padded, encrypted and afterwards the formatted encrypted data is input to the authentication calculation. Currently, cf. \cite{ICAO11}, four combinations are specified:
\begin{description}
\item[id-CA-DH-3DES-CBC-CBC] This uses two key 3DES \cite{3DES} in Cipher Block Chaining (CBC) mode for both encryption and authentication (112 bit security).
\item[id-CA-DH-AES-CBC-CMAC-128] This uses 128 bit AES \cite{AES} in Cipher Block Chaining (CBC) mode for encryption and in CMAC mode for authentication (128 bit security).
\item[id-CA-DH-AES-CBC-CMAC-192] This uses 192 bit AES \cite{AES} in Cipher Block Chaining (CBC) mode for encryption and in CMAC mode for authentication (192 bit security).
\item[id-CA-DH-AES-CBC-CMAC-256] This uses 256 bit AES \cite{AES} in Cipher Block Chaining (CBC) mode for encryption and in CMAC mode for authentication (256 bit security).
\end{description}
Although in principle an e-passport might support multiple algorithms, in practice an e-passport only supports one.  In Protocol \ref{CA} we have given a description based on \cite{ICAO11} of how a party (terminal) can run the CA protocol.  The terminal can only conclude a CA execution is successful, if it successfully read a file on the document over the established secure messaging. This is why the protocol \ref{CA} also contains a file identifier $F_\text{Id}$ and a (non-zero) number of bytes to be read.
There exist two methods to read a file on an e-passport (cf. \cite[Section 3.9.3]{ICAO10}): by selecting the file and then reading the data, or by reading the data directly using a so-called short file identifier. The first method uses two commands. Compare \cite{ISO/IEC7816-4}. The second and recommended method only uses one such command called READ BINARY (RB) on which we have based Protocol \ref{CA} on.
\vspace{-3mm}
\begin{algorithm}[H]
\floatname{algorithm}{Protocol}
\caption{\em CA protocol and reading $n$ bytes from file $F_\text{Id}$ with command $RB$.}\label{CA}
  \begin{algorithmic}[1]
   \Statex \texttt{[{\em We assume the terminal has successfully setup a BAC/PACE channel.}]}
   \State \texttt{The terminal reads data group 14 (CA public key $Y$, domain parameters and encryption algorithms to use) from the e-passport.}
   \State \texttt{The terminal validates the authenticity of the data group through passive authentication. If this fails, CA fails.}
  \State \texttt{The terminal generates an {\em ephemeral} Diffie-Hellman key pair, i.e. generates a (pseudo) random number $k \in (0, q)$ and forms the ephemeral public key $Z=kG$. This is sent to the e-passport.}
  \State \texttt{Both parties compute the shared Diffie-Hellman secret $K$. The card calculates this as $K=xZ$, the terminal as $K=kY$. Both parties derive two session keys $\text{KS}_\text{Enc}$ and $\text{KS}_\text{MAC}$ from their view of $K$.}
  \State \texttt{On these keys, CA secure messaging will be based, encryption is based on key $\text{KS}_\text{Enc}$ and authentication on key $\text{KS}_\text{MAC}$.}
  \State \texttt{The terminal sends the protected command $\overline{RB}$ to read $n$ bytes from file $F_\text{Id}$ resulting in a protected response ${\bar M}$, i.e. ciphertext.}
  \State \texttt{If cryptographic processing (authentication validation, decryption) on ${\bar M}$ fails, CA fails otherwise CA is successfully executed.}
  \end{algorithmic}
  \end{algorithm}
\vspace{-6mm}
Successful secure messaging (Steps 6,7) is a guarantee that the two views of $K$ in Step 4 coincide and that the e-passport is authentic. In normal practice a terminal application never uses the protected ${\bar M}$ but will see the actual plaintext bytes read from the file and a successful return status. The following proposition is the basis for the working of our RDE implementation; we let $|q|$ denote the size in bits of the CA group order.

\begin{proposition} \label{prop:RDE-key}
The protected messages $\overline{RB}$ and ${\bar M}$ in Step 6 of Protocol \ref{CA} are deterministically based on the ephemeral key $Z$ from Step 3, the file $F_\text{Id}$ read and the actual contents read. That is, if the terminal executes Protocol \ref{CA} twice with the same ephemeral key $Z$ and the same parameters $(n, F_\text{Id})$, then $\overline{RB}$ and ${\bar M}$ would also be the same.
\end{proposition}
{\bf Proof:}
We only discuss ${\bar M}$ as results on $\overline{RB}$ follow similarly.
The proof of the first part of the proposition follows from an analysis of the CA secure messaging setup in \cite{ICAO11}.
If follows from \cite[Section 9.7]{ICAO11} that the derived keys $\text{KS}_\text{Enc}$ and $\text{KS}_\text{MAC}$ are deterministically based on $K$ only. Moreover, $K$ is deterministically based on the ephemeral key $Z$ and the private CA key of the e-passport. Secure messages either contain commands from the terminal to the card, or responses from the card. It follows from \cite[Section 9.8]{ICAO11} that protected messages contain an encrypted message and a ``cryptographic checksum''. The message to be encrypted (card command or response) is first padded in a deterministic fashion (\cite[``Padding Method 2'']{ISO/IEC9797-1} ) and then encrypted with $\text{KS}_\text{Enc}$. It follows that the encrypted messages deterministically follow from the plain input (card command or response) and $\text{KS}_\text{Enc}$ (and hence $Z$). If also follows from \cite[Section 9.8]{ICAO11} that the cryptographic checksum is a message authentication code based on $\text{KS}_\text{MAC}$ on the encrypted message and a so-called send sequence counter (SSC). This SSC starts with value zero and is increased with each command and response. That is, the read command of Step 7 of Protocol \ref{CA}, first uses SSC value 1 is used and the card response uses SSC value 2 etcetera. It follows that the cryptographic checksums also deterministically follows from $K$, the plain input (card command or response) and SSC, i.e. the number of the protected message.
\hfill $\Box$

The previous results implies that the protected response ${\bar M}$ in Protocol \ref{CA} is a function of $n, F_\text{Id}$ and the (pseudo)random $k$, which we represent by ${\bar M}(.,.,.)$. That is, ${\bar M} = {\bar M}(n, F_\text{Id}, k)$. Similarly we specify a function $\overline{RB}(n, F_\text{Id}, k)$.  We remark that it is crucial that in CA messaging the send sequence counter (SSC) starts with one and is not randomly set.
The latter is the case in BAC messaging when based on 3DES, cf. \cite[Section 9.8.6.3]{ICAO11}. As CA messaging completely replaces BAC messaging this is of no consequence. In addition to the analysis of the e-passport specification \cite{ICAO11} we also successfully tested the first part of Proposition \ref{prop:RDE-key} using the Smart Card Shell and the Smart Card Shell Script Collection \cite{SCS}. Our tests were based on the e-passport simulator in the collection and on a Dutch e-passport. In the simulation the test was performed with 3DES and 128 bit AES. The CA protocol of the Dutch e-passport was based on the Brainpool320r1 curve and 256 bit AES secure messaging. In the appendix we have provided results from the simulator tests.
The following assumption is the basis for the security of our RDE implementation.

\begin{assumption} \label{assumption}
Let $n, F_\text{Id}$ be fixed.
Then the functions $k \rightarrow \overline{RB}(n, F_\text{Id}, k)$ and $k \rightarrow {\bar M}(n, F_\text{Id}, k)$ are key derivation functions sensu \cite{NISTSP800108} of security strength at least $\tt{MIN}(128, |q|/2 )$ bits for DES based secure messaging and $\tt{MIN}(192, |q|/2 )$ bits for AES.
\end{assumption}

Effectively this assumption states that the ICAO secure messaging mechanism \cite{ICAO11} is sound: 
to break the mechanism an attacker is forced to either find the ephemeral Diffie-Hellman private key $k$ or to brute force 
$\text{KS}_\text{Enc}$ and/or $\text{KS}_\text{MAC}$. Eavesdropping on communication between terminal and e-passport and partial plaintext knowledge should not help an attacker effectively to extract plaintext from new secure messages or in predicting them.

This assumption can be further quantified by inspecting the secure messaging setup used in e-passports, cf. \cite[Figure 5]{ICAO11}.
A protected message consists of at least one full encryption block based on $\text{KS}_\text{Enc}$ and
of a 64 bit message authentication code over the encrypted payload based on $\text{KS}_\text{MAC}$.
It is reasonable to assume that each such cryptogram behave as the output of a key derivation function sensu \cite{NISTSP800108} based on the encryption key used.
Actually, \cite{NISTSP800108} specifies key derivation functions in such fashion.
In case of DES based messaging the encryption block is of size 64 bits and in case of AES this is of size 128 bits.
So the total length is at least 128 bits in case of DES based messaging and at least 192 bits in case of AES.
Both encryptions keys $\text{KS}_\text{Enc}$ and $\text{KS}_\text{MAC}$ are derived from the shared Diffie-Hellman key based on $k$ .
The final motivation for the assumption is that the total security strength of the considered function is upper bounded by the security strength of the CA protocol which is $|q|/2$ bits, cf. \cite{KATZ}.

\section{RDE specification}\label{sec:RDE_spec}
Based on the discussions in Section \ref{sec:CA_and_RDE} we can now specify Remote Document Encryption.
The idea is use the protected response ${\bar M}$ from Protocol \ref{CA} , i.e. a cryptogram, to derive an encryption key from.
A party can compute ${\bar M}$ by simulation and the e-passport holder can use its passport.
Compare Figure \ref{fig:rde_exp} from the appendix: the first rounds corresponds with the party and the second round with the user.

RDE consists of three use cases: e-passport holder registration at a party, data encryption by the party and decryption of the data by the holder using his e-passport. These uses cases are specified in Protocols \ref{RDE_registration}, \ref{RDE_ENC} and \ref{RDE_DEC} respectively. These protocols are examples; some steps can be done in different order too. In these protocols we introduce the notion {\em RDE key extraction parameters}. This is a triple $(n, F_\text{Id}, F_\text{Cont})$ where the first two elements are from Protocol \ref{CA}. We let $F_\text{Id}$ be a PA protected data group  and we let $F_\text{Cont}$ represent the full contents of it. In practice it is convenient to let $F_\text{Id}$ simply be data group 14, i.e. the data group holding the CA public key.
\begin{algorithm}[H]
\floatname{algorithm}{Protocol}
\caption{\em Remote Document Encryption registration}\label{RDE_registration}
\begin{algorithmic}[1]
\State \texttt{The e-passport holder authenticates to a party wanting to use RDE. If authentication fails, RDE registration fails.}
\State \texttt{The e-passport provides the party with data group 1 (Identity Information), data group 14 (CA public key and parameters) and the Document Security Object  ($\text{EF.SO}_\text{D}$) from its e-passport.}
\State \texttt{The party validates consistency of the information given in Steps 1 and 2. The authenticity of the information provided in Step 2 is validated through passive authentication. If either validation is unsuccessful, RDE registration fails.}
\State \texttt{The holder and the party agree on RDE key extraction parameters $(n, F_\text{Id}, F_\text{Cont})$. Here the party validates the authenticity of $F_\text{Cont}$ through passive authentication using $\text{EF.SO}_\text{D}$ from Step 2.}
\State \texttt{The holder gives his consent to the party for the usage of RDE to encrypt (confidential) information. If consent is not given, RDE registration fails.}
\State \texttt{The holder is registered at the party for the usage of RDE.}
\end{algorithmic}
\end{algorithm}
\vspace{-6mm}
Certainly the first two steps of RDE registration could be based on, and in fact combined with, RDA.
As indicated earlier, for convenience one can let data group 14 also take the role of file $F_\text{Id}$.
We note that like all cards, e-passports have certain restrictions on the size a command plus parameters and a response can take.
All e-passports support a standard size response of 255 bytes. With secure messaging extra overhead is introduced limiting the effective data one can read. Typically

In the next protocol we specify
how the party can encrypt for the e-passport holder with the registered data. Here $H(.)$ represents a secure hash function, e.g. SHA-256 \cite{SHA256}. We let $\ENC_K (.)$ represent an symmetric encryption operation, e.g. AES \cite{AES}.
\vspace{-4mm}
\begin{algorithm}[H]
\floatname{algorithm}{Protocol}
\caption{\em Remote Document Encryption of data $D$ for e-passport holder $H$}\label{RDE_ENC}
\begin{algorithmic}[1]
\State \texttt{The party looks up the CA public key of the holder and the RDE key extraction method $(n, F_\text{Id}, F_\text{Cont})$.}
\State \texttt{The party generates a ephemeral Diffie-Hellman key pair and uses this together with the CA public key to calculate session keys $\text{KS}_\text{Enc}$ and $\text{KS}_\text{MAC}$ from CA secure messaging.}
\State \texttt{The party simulates reading the holder's e-passport in line with extraction method $(n, F_\text{Id}, F_\text{Cont})$ resulting in a protected READ BINARY command $\overline{RB}$ and protected response ${\bar M}$.}
\State \texttt{The party uses the protected response ${\bar M}$ as the output of a key derivation function, i.e. uses a secure hash value $H({\bar M})$ of ${\bar M}$ to encrypt data $D$, i.e. forms $C =\ENC_{H({\bar M})} (D)$.}
\State \texttt{The party provides the ephemeral public key $Z$, $\overline{RB}$ and $C$ to the holder.}
\end{algorithmic}
\end{algorithm}
\vspace{-6mm}
In the next protocol we specify how the holder can decrypt the encrypted data $C$ by using his e-passport.
\begin{algorithm}[H]
\floatname{algorithm}{Protocol}
\caption{\em Remote Document Decryption of $Z, \overline{RB}, C$ by e-passport holder $H$}\label{RDE_DEC}
\begin{algorithmic}[1]
\State \texttt{The holder uses $Z$ this together with his e-passport to start CA secure messaging.}
\State \texttt{The holder sends the protected READ BINARY command $\overline{RB}$ to his e-passport,
 resulting in protected response ${\bar M}$.}
\State \texttt{The holder uses the secure hash value $H({\bar M})$ of ${\bar M}$ to decrypt $C$.}
\end{algorithmic}
\end{algorithm}
\vspace{-6mm}
RDE security follows from Assumption \ref{assumption}.
That the protected READ BINARY command $\overline{RB}$ will not help an attacker to predict 
the protected response ${\bar M}$, is closely related to soundness of ICAO secure messaging \cite{ICAO11}.

It is best to base $\ENC_K (.)$ on an authenticated form of encryption, e.g. AES based on CCM \cite{CCM}.
By doing so the holder has assurance the RDE decryption process was conducted correctly.
One can omit the second step (authentication) in Protocol \ref{RDE_registration}, but this introduces the risk that a fraudster uses a stolen e-passport to register the holder and gets access to confidential information. This can be mitigated by adding holder authentication in Step 5 of Protocol \ref{RDE_ENC}. In the Secure Messagebox and Medical Portal Compartmentalization use cases from Section \ref{sec:intro} one can have holder authentication in both Protocols \ref{RDE_registration} and \ref{RDE_ENC}.

We have experimented with RDE on an AES-256 based Dutch passport (issued June 2015, based on Brainpool320r1/AES-256) and a Dutch identity card (issued July 2010, based on Brainpool256r1/3DES). In the first case one achieves 160 bit security and in the second 128 bit, cf. Assumption \ref{assumption}.
In both cases the RDE decryption execution time on the e-passport/card was less than 2 seconds.
\section{Further discussions, extensions and variations}\label{sec:extensions}
According to NIST \cite{NISTSP80057}, 128 bit security suffices beyond 2031. To estimate the strength of 160 bit security achievable by e-passports we use Moore's law. A widely accepted interpretation of this law is that that the computing power one gets for the same amount of money doubles every 18 months. This means that 160 bit security would suffice beyond the year 2079 (=2031 +  32*1,5). Our results ironically suggest that carrying a passport when traveling abroad might violate export or import laws on cryptography. Compare \cite{Koops}.

In many countries, including the Netherlands, the national identity card is also based on the ICAO specifications \cite{ICAO}.
Most notably, the fifth part of these specifications define smartcard sized (``TD1'') documents corresponding to typical identity cards.
For such national identity cards, RDE is also available if CA is supported. We successfully verified that a Dutch identity card (issued in July 2010) conforms to Proposition \ref{prop:RDE-key}. RDE can also be available on electronic driver licenses conformant to the European regulation 383/2012
of 4 May 2012 \cite{EU383/2012} and/or standard ISO/IEC 18013 \cite{ISO/IEC18013} as both have adopted the ICAO CA protocol. Electronic driver licenses are in use in the Netherlands since 14 November 2014. These support the BAC/AA protocol but not the CA protocol implying that RDE is not possible on them. RDE seems not possible on the German eID card ({\em neue Personalausweis}) \cite{BSI2}. A first reason for this is the switching of the order of CA and TA execution(CA/TA version 2). A terminal first has to successfully conclude the TAv2 protocol, i.e. has to be authenticated, before the CAv1 protocol can start. A more fundamental issue is that unlike in e-passport CA CAv2 secure messaging is based on a random number generated by the card. This makes it impossible for a party to predict protected responses of the card.

In Protocol \ref{CA} we only use one protected READ BINARY to base an RDE encryption key on. Alternatively, one could use a few of such commands and base the encryption key on all secure responses. By using $i$ read commands of, for instance, the first $i$ bytes of data group 14, the $i$-th security response would contain a 64-bit authenticator also based on send sequence counter 2, 4, \ldots , $2\cdot i$. In this way the encryption key closely resembles the NIST key derivation function ``KDF in Counter Mode'' \cite[Section 5.1]{NISTSP800108}. Note that one would then need $i$ protected READ BINARY commands corresponding to send sequence counter 1, 2, \ldots , $2\cdot i -1$
To harvest this conceptual advantage, it typically suffices to take $i=2$.

We can easily supplement the RDE possession factor with a knowledge factor by introduction
of the notion of RDE ``Extraction PIN'' (or ``Extraction Password'').
This PIN is then agreed between the holder and the party wanting to use RDE as part of
of Step 4 of RDE registration, i.e. Protocol \ref{RDE_registration}. The party then uses this PIN to additionally encrypt
the elements $Z$ and/or $\overline{RB}$ in Step 5 of Protocol \ref{RDE_ENC} in such a way that the PIN cannot be deduced from the encrypted elements only, i.e. without the e-passport. As an illustration, simply AES encrypting $Z$ with the PIN would not be sound. Indeed, an attacker could find the PIN by looking for a PIN that decrypts the ciphertext to an point in $\langle G \rangle$. A simple and effective way to do this is to deterministically map the PIN in the group $\langle G \rangle$ used in the Chip Authentication protocol and then add the embedded PIN to $Z$. That is, all parties agree upon an embedding $\EMB: \{0,1\}^* \rightarrow \langle G \rangle$ and
the encrypting party sends $Z'= \EMB ( \mbox{PIN} ) + Z$ instead of $Z$ to the holder in Step 5 of Protocol \ref{RDE_ENC} .
As part of the RDE decryption the holder uses his PIN to calculate $Z' - \EMB ( \mbox{PIN} )$ and uses this as ephemeral key $Z$. As $Z$ is a random point so is $Z'$, implying that it holds no information on the PIN. We note that \cite{ICAO11} specifies two suitable mappings (generic and integrated). One can also consider the PIN as a big integer and look for a point in $\langle G \rangle$ with that PIN as its x coordinate. If this fails one add one to the PIN until such a point exists. In commonly used elliptic curves two such points exists corresponding with a y coordinates that are even and odd. By consistently choosing one such point one obtains a deterministic mapping. Many variants of this construction exist. An RDE extraction PIN is not a regular one in the sense that a passport does not lock after a number of incorrect PINs are used by the holder. However, an RDE extraction PIN still substantially reduces the risk related to loss of the passport by the legitimate holder. Indeed, with possession of the passport and RDE encrypted data but without the Extraction PIN a party would need to brute force the Extraction PIN in {\em interaction} with the passport. That is, the RDE extraction PIN is bound to the e-passport which is an important property.
Each try takes about 1 second. This implies for instance that the brute force approach has an expected run time of more than 17 years for a 5 digit PIN from an alphabet of 64 characters. Indeed as  $64^5 / (2*3600*24*365) \approx 17$. Actually, it seems unlikely that a passport will be be able to withstand so many operations without failure.

\section{Conclusion}\label{sec:conclusion}
Based on the Chip Authentication protocol we have demonstrated Remote Document Encryption (RDE) for e-passports, identity cards and electronic driver licenses.
This allows any party to encrypt data for an electronic document holder such that only with possession of the document these can be decrypted.
Possible applications include additional protection of personal data stored in various types of web applications and protection of local storage in mobile applications.
Combined with Remote Document Authentication \cite{RDA} one can effectively transform an e-passport into a complete PKI smartcard albeit not PIN protected in the regular sense. To this end we have introduced the notion of Extraction PIN, effectively providing the same security.
We finally remark that RDE encrypted data is lost when the document used is lost.
This is particulary relevant as in some countries old identity documents are taken as part of renewal.

\section{Acknowledgement}
We are indebted to Gert Maneschijn for posing the question on feasibility of Remote Document Encryption on e-passports and its applications. We want to thank Andreas Schwier for modifying the eID simulator script collection to let it support AES-128 CA secure messaging. Finally we want to thank Alfred Velthuis for initially pointing us to the possibilities of the Smart Card Shell and Smart Card Shell Script Collection.

\newpage
\appendix
\section{RDE experiments} \label{sec:simulator_tests}
\begin{figure}[H]
\includegraphics[scale=0.47,center]{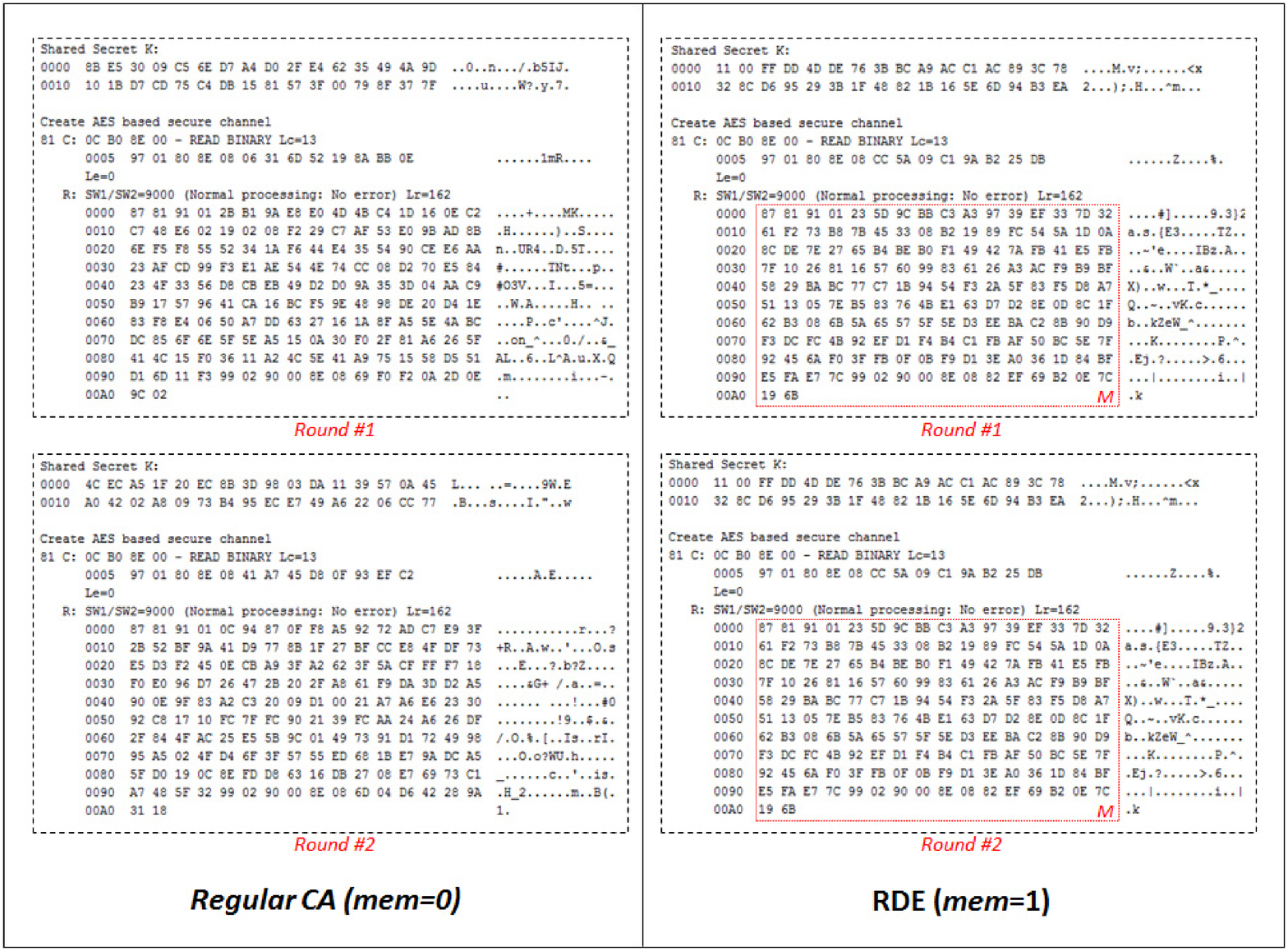}
\caption{RDE experiments}\label{fig:rde_exp}
\end{figure}

We have tested Proposition \ref{prop:RDE-key} in practice using the Smart Card Shell and the Smart Card Shell Script Collection \cite{SCS}. For this we slightly modified the method {\em generateEphemeralCAKeyPair} in the JavaScript ``chipauthentication.js'' in the ICAO directory of the collection. The modified method copies the ephemeral key pair (prkCA and pukCA) and reuses this in the next call of the method if a variable $mem$ is set to one. We also introduced a modification of the method {\em readEFwithSFI} JavaScript ``eac20.js''. The original method keeps reading the elementary file using the READ BINARY command until an error is encountered. This new method, {\em read128EFwithSFI}, only uses one READ BINARY command reading (maximally) 128 bytes from the file typically without returning an error. In our experiment we used the Smart Card Shell JavaScript code indicated below. We think this script is self-explanatory but for further explanation we refer to \cite{SCS}.
In the script we read data group 14 holding the CA key as suggested after Protocol \ref{RDE_registration}.
The MRZ in the script below allows the script to work out-of-the box with the e-passport simulator in the Smart Card Shell Script Collection.
\begin{Verbatim}[frame=single]
load("eac20.js");
var mrz = "TPD<<T220001293<<<<<<<<<<<<<<<6408125<1010318D"+
          "<<<<<<<<<<<<<6MUSTERMANN<<ERIKA<<<<<<<<<<<<<";
var card = new Card(_scsh3.reader);
card.reset(Card.RESET_COLD);
/* Round 1 */
card.reset(Card.RESET_COLD);
var crypto = new Crypto();
var eac = new EAC20(crypto, card);
eac.selectLDS();
eac.performBACWithMRZ(mrz);
eac.readDG14();
eac.readCVCA();
eac.performChipAuthentication();
eac.read128EFwithSFI(14);
/* Round 2 */
mem=1; /* reuses ephemeral key pair */
       /* mem=0 generates new key pair */
card.reset(Card.RESET_COLD);
var crypto = new Crypto();
var eac = new EAC20(crypto, card);
eac.selectLDS();
eac.performBACWithMRZ(mrz);
eac.readDG14();
eac.readCVCA();
eac.performChipAuthentication();
eac.read128EFwithSFI(14);
\end{Verbatim}
When this script is run in the Smart Card Shell it displays the shared key, the secure messages and the decrypted messages.
In Figure \ref{fig:rde_exp} we have depicted the shared keys and secure responses of the final call {\em eac.read128EFwithSFI(14)} in the script.
Here the Smart Card Shell simulator was used to simulate an AES-128 based e-passport.
The left side depicts the two protected responses in the regular situation ($mem=0$): the shared keys and responses from rounds 1 and 2 are different.
The right side corresponds with the RDE setting ($mem=1$): here the shared keys and protected responses (i.e. $\cal M$ from Proposition \ref{prop:RDE-key}) from rounds 1 and 2 coincide.
This behaviour is in line with Proposition \ref{prop:RDE-key}.
Similar experiments were done with a simulation of a 3DES based e-passport.
By changing the MRZ in the script we conducted an experiment on an AES-256 based Dutch passport (issued June 2015, CA based on Brainpool320r1/AES-256).
We also tested a Dutch identity card (issued July 2010, CA based on Brainpool256r1/3DES).
All experiments are consistent with Proposition \ref{prop:RDE-key}.

\end{document}